\documentclass[nonblindrev]{informs3} 

\usepackage{balance}
\usepackage{amsmath}
\usepackage[tight,footnotesize]{subfigure}
\usepackage{graphicx}
\usepackage{bm}
\usepackage{algorithm}
\usepackage[noend]{algorithmic}
\usepackage{multirow}
\usepackage{slashbox}
\usepackage{caption}
\usepackage{amsfonts}

\OneAndAHalfSpacedXII 

\usepackage{natbib}
 \bibpunct[, ]{(}{)}{,}{a}{}{,}%
 \def\bibfont{\small}%
\TheoremsNumberedThrough     

\EquationsNumberedThrough    

\MANUSCRIPTNO{2017} 

\begin{document}


\RUNAUTHOR{Yuan et al.}

\RUNTITLE{No Time to Observe: Adaptive Influence Maximization with Partial Feedback}

\TITLE{No Time to Observe: Adaptive Influence Maximization with Partial Feedback}

\ARTICLEAUTHORS{%
\AUTHOR{Jing Yuan Shaojie Tang}
\AFF{The University of Texas at Dallas}
} 

\ABSTRACT{    Although influence maximization problem has been extensively studied over the past ten years, majority of existing work adopt one of the following models: \emph{full-feedback model} or \emph{zero-feedback model}. In the zero-feedback model, we have to commit the seed users all at once in advance, this strategy is also known as non-adaptive policy. In the full-feedback model, we select one seed at a time and wait until the diffusion completes, before selecting the next seed. Full-feedback model has better performance but potentially huge delay, zero-feedback model has zero delay but poorer performance since it does not utilize the observation that may be made during the seeding process. To fill the gap between these two models, we propose \emph{Partial-feedback Model}, which allows us to select a seed at any intermediate stage. We develop two novel greedy policies that, for the first time, achieve bounded approximation ratios under both uniform and non-uniform cost settings. }


\maketitle

%

\section{Introduction}\label{sec:introduction}

Since the seminal work of \citep{domingos2001mining}, the influence maximization problem has attracted tremendous attention in recent years. This problem is first formalized and studied by \citep{kempe2003maximizing} as a discrete optimization problem. They study this problem under several diffusion models including independent cascade model and linear threshold model. They demonstrate that the influence maximization problem under both models are NP-hard, however, the objective function is monotone and submodular. Leveraging these nice properties, they propose an elegant greedy algorithm with constant approximation ratio.  Since then, considerable work  \citep{chen2013information,leskovec2007cost,cohen2014sketch,chen2010scalable,chen2009efficient,tang2011relationship,tang2016optimizing,tong2016adaptive,yuan2017discount} has been devoted to this topic and its variants.

However, majority of existing work fall into one of the following categories: \emph{full-feedback model} \citep{golovin2011adaptive} or \emph{zero-feedback model} \citep{kempe2003maximizing}. In the zero-feedback model, we have to commit the seed users all at once in advance, this strategy is also known as non-adaptive policy. 
In the full-feedback model, we select one seed at a time and wait until the diffusion completes, before selecting the next seed, this policy is also known as adaptive policy. This type of policy has better performance in terms of  expected cascade because of its adaptivity, e.g, it allows us to adaptively choose the next seed after observing the actual spread resulting from previously selected seeds. However, the viral marketing in reality is often time-critical, implying that it is impractical, sometimes impossible,  to collect the full status of the actual spread before selecting the next seed.

 To fill this gap, we propose a generalized feedback model, called \emph{partial-feedback model}, that captures the tradeoff between performance and delay. We adopt \emph{independent cascade model} (IC) \citep{kempe2003maximizing}, which is one of the most commonly used models, to model the diffusion dynamics in a social network. Unfortunately, we show that the objective under partial-feedback model and IC is not adaptive submodular, implying that existing results on adaptive submodular maximization does not apply to our model directly. To overcome this challenge, we develop two novel greedy policies, that, for the first time, achieve bounded approximation ratios under partial-feedback model. One nice feature of our policies  is that we can balance the delay/performance tradeoff by adjusting the value of a controlling parameter. 

\subsection{Related Work}
\label{sec:related}
In \citep{domingos2001mining}, they show that data mining can be used to determine potential seed users in viral market. Since then, there is a rich body of works that has been devoted to viral marketing problem.  Most of existing works on this topic can be classified into two categories.

The first category is non-adaptive influence maximization: we must find a set of influential customers all at once in advance subject to a budget constraint. Kempe et al. \citep{kempe2003maximizing} first formalized and studied this problem under two diffusion models, namely Independent Cascade model and Linear Threshold model.  \citep{chen2013information,leskovec2007cost,cohen2014sketch,chen2010scalable,chen2009efficient}  study influence maximization problem under various extended models. 

The second category is adaptive influence maximization, which is closely related to adaptive/stochastic submodular maximization \cite{golovin2011adaptive,badanidiyuru2016locally,tong2016adaptive,yuan2017discount}. Existing studies mainly adopt full-feedback model, assuming that we can observe the full status of the previous cascade before selecting the next seed. We relax this assumption by  incorporating partial-feedback and develop two  adaptive policies that achieve the first bounded approximation ratios under partial-feedback model.
\section{Network Model and Diffusion Process}
\label{sec:system}
\subsection{Independent Cascade Model} A social network is modeled as a directed graph  $G = (V, E)$, where $V$ is a set of $n$ nodes
and $E$ is a set of social ties.  We adopt \emph{independent cascade model} \citep{kempe2003maximizing} to model the diffusion dynamics in a social network. Each node $v\in V$ is associated with a cost $c_v$, each edge $(u, v)$ in the graph is associated with a propagation probability $p_{uv}$, which is the probability that node $u$ independently influences node $v$ in the next  slot after $u$ is influenced. 
The expected cascade of $U$, which is the expected number of influenced nodes given seed set $U$, is denoted as $I(U)$.

\subsection{The Feedback Model}
Majority of existing work adopt one of the following feedback models: \emph{zero-feedback model} or \emph{full-feedback model}.

 \emph{\underline{Zero-feedback Model:}} During the seeding process, we can not observe anything about the resulting spread of adoption. Since there is no benefit in ``waiting'', we can simply commit the seeds all at once in advance. This model is equivalent to traditional  non-adaptive influence maximization problem which has been well studied in the literature, \citep{kempe2003maximizing} showed that the utility function under this model is submodular and then propose a Greedy algorithm whose performance is lower bounded $(1-1/e)$ times the optimal non-adaptive solution. It is still unclear about the upper bound on the adaptivity gap, e.g., the performance gap between non-adaptive and adaptive solutions.

\emph{\underline{Full-feedback Model:}}  We select one seed at a time and wait until the diffusion completes, before selecting the next seed, this policy is also known as adaptive policy. In particular, after selecting a seed $u$, we can observe the status of all edges existing $v$, where $v$ is any node that is reachable from $u$ via live edges in the full realization. For this model, \citep{golovin2011adaptive}  introduced the concept of adaptive submodularity and proposed  a Greedy algorithm has a $(1-1/e)$ approximation guarantee.

As discussed in Section \ref{sec:introduction}, full-feedback (resp. zero-feedback) model has better (resp. poorer) performance but potentially huge (resp. small) delay.  To fill the gap between these two models, we propose \emph{partial-feedback model}, which generalizes the previous two models by allowing us to select the next seed at some intermediate stage.

Based on independent cascade model, we next introduce the concept of \emph{diffusion realization} \citep{golovin2011adaptive} and other related concepts.
\begin{definition}[Diffusion Realization]  Let $\psi((u,v))\in\{0,1,?\}$ denote the state of $(u,v)$ given realization $\psi$ where $\psi((u,v))=0$ means that $(u,v)$ is blocked, $\psi((u,v))=1$ means that $(u,v)$ is live, and $\psi((u,v))=?$ means that the state of $(u,v)$ is unknown.
We represent the state of the diffusion stage using function $\psi: E \rightarrow \{0,1, ?\}$, called diffusion realization.
\end{definition}
\begin{definition}[Partial and Full Realization]  Let $\mathrm{dom}(\psi) = \{(u,v) : \psi((u,v))\in\{0,1\}\}$ denote  the set of edges observed in $\psi$. $\psi$ is a full realization if  $\mathrm{dom}(\psi)=E$ (i.e., all edges are observed in full realization). Otherwise, $\psi$ is called partial realization.
\end{definition}
 \begin{definition}[Intermediate and Final Realization]  Given a realization $\psi$ of seeds $S$, we say $\psi$ is a final realization if $\mathrm{dom}(\psi)$ is finalized (i.e., we can not observe more edges other than  $\mathrm{dom}(\psi)$ by waiting longer).  Otherwise, $\psi$ is called intermediate realization. Let $X(S)$ denote the set of all final realizations of $S$.
\end{definition}
 \begin{definition}[Subrealization and Superrealization]  A  realization $\psi_a$ is consistent with a realization $\psi_b$ (i.e., $\psi_a \thicksim \psi_b$) if they are equal everywhere in $\mathrm{dom}(\psi_a)\cap \mathrm{dom}(\psi_b)$. We say $\psi_a$ is a subrealization of $\psi_b$ (or equivalently $\psi_b$ is a suprealization of $\psi_a$) if $\psi_a$ is consistent with  $\psi_b$ and $\mathrm{dom}(\psi_a) \subseteq\mathrm{dom}(\psi_b)$, or $\psi_a \subseteq \psi_b$. We say $\psi_b$ is a final superrealization of $\psi_a$ given $S$ if $\psi_b \in X(S)$ and  $\psi_b\supseteq   \psi_a$ (i.e., $\psi_b$  is a final realization and $\psi_b$ is a superrealization of $\psi_a$).
\end{definition}

 Consider an intermediate realization $\psi$ of seeds $S$, let $X_{\mathrm{final}}(S|\psi):=\{\phi:\phi \in X(S), \phi\supseteq  \psi\}$ denote the set of all final realizations of $S$ given $\psi$.  


\section{Problem Formulation}
Under the partial-feedback model, we perform the decision process in a sequential manner where the decision made in each round is depending on the current observation of the network status and the remaining budget.

\begin{definition}[Adaptive Policy]
 We define our adaptive policy $\pi: (\psi, S) \rightarrow v$, which is a function from the current ``observation'' $\psi$ and the set of existing seeds $S$ to $v$, specifying which seed to pick next given a particular set of observations.
 \end{definition}
 \begin{definition}[Policy Concatenation] Given two policies $\pi_a$ and $\pi_b$, we use  $\pi_a @ \pi_B$ to denote a policy
that runs $\pi_a$ first, and then runs policy $\pi_b$, ignoring the information obtained from running $\pi_a$.
 \end{definition}

 We use $\Psi$ to denote a random realization. Assume there is a known prior probability distribution  $\Pr[\psi] := \Pr[\Psi = \psi]$  over $W$ which is the universe  of all possible full realizations.  Given a full realization $\psi$, let $S_{\psi}$ denote all seeds picked by $\pi$ under $\psi$, and $c(S_{\psi})=\sum_{v\in S_{\psi}} c_v$ denote the total cost of $S_{\psi}$. The expected cascade of a policy $\pi$ is
 \[f(\pi) = \mathbb{E}[I(S_{\Psi}| \Psi)] = \sum_{\psi\in W} \Pr[\psi] I(S_{\psi}) \]
  where $I(S_{\psi})$ denotes the cascade of $S_{\psi}$ given realization $\psi$, e.g., all nodes that are reachable from $S_{\psi}$ under $\psi$. The goal of the adaptive influence maximization problem is to find a policy $\pi$ such that
\begin{center}
\framebox[0.45\textwidth][c]{
\enspace
\begin{minipage}[t]{0.45\textwidth}
\small
\emph{Maximize $f(\pi)$}\\
\textbf{subject to:}
\begin{equation*}
c(S_{\psi}) \leq B, \forall  \psi\in W
\end{equation*}
\end{minipage}
}
\end{center}
\vspace{0.1in}

We first show that the objective is not adaptive submodular under partial feedback model. The concept of adaptive submodularity is a generalization of submodularity to adaptive policies: we say a function $f$ is adaptive submodular if adding an element $e$ to a realization $\psi$ increases $f$ at least as much as adding $e$ to a superset of $\psi$. Since the Myopic Feedback model proposed in \citep{golovin2011adaptive} is a special case of our partial-feedback model, we borrow the same counter example from their work to prove the following lemma.
\begin{theorem}\citep{golovin2011adaptive}
The objective $f$ under partial-feedback model is not adaptive submodular.
\end{theorem}
 First of all, we want to emphasize the difference between ``round'' and ``slot''. One round corresponds to one execution of our algorithm, while one slot corresponds to one step of information propagation. In the rest of this paper, we use $|\cdot|$ to denote the size of a set.
\section{$\alpha$-Greedy  Adaptive Policy with Partial Feedback}
In this section, we present the first adaptive policy, called $\alpha$-Greedy policy, and analyze its performance bound. For ease of presentation, we start with uniform cost setting where all nodes have identical costs. Then we extend this result to non-uniform cost setting.
\subsection{Uniform Cost}
We first study the case with uniform cost, e.g., $\forall v\in V: c_v=1$. Since each node has the same cost, the budget constraint can be reduced to cardinality constraint, e.g., the number of seeds that can be selected is upper bounded by $B$. 

\subsubsection{Policy Description}
$\alpha$-Greedy policy $\pi^\alpha$ (Algorithm \ref{alg:LPP1}) is performed in a sequential greedy manner as follows: After observing the current partial realization, we choose to either wait for another slot or select the next seed immediately that maximizes the expected marginal benefit. This process iterates until  the budget is used up.  
\begin{enumerate}
\item Start with slot $t=1$, seeds $S=\emptyset$, and a control parameter $\alpha\in[0,1]$;
\item Suppose we have made observations $\psi_{[t]}$ at slot $t$,  let $p_v(S; \psi_{[t]})$ denote the activation probability of $v$ given seeds $S$ and observation $\psi_{[t]}$, and let $f(S| \psi_{[t]})=\sum_{v\in V} p_v(S; \psi_{[t]})$ denote the expected cascade under the same setting. Define ${O(S; \psi_{[t]})}:=\{v: p_v(S; \psi_{[t]})=0\}$ as the set of nodes whose activation probability is zero under $\psi_{[t]}$ and $S$.  We then examine the following condition.
        \begin{equation}
\label{cond:!}
\frac{f(S| \psi_{[t]})}{|V\setminus O(S; \psi_{[t]})|}\geq \alpha
 \end{equation}
 If  condition (\ref{cond:!}) is satisfied,  select $\arg\max_{u \in V }\Delta_u(S; \psi_{[t]})$ where  $\Delta_u( \psi_{[t]}; S)= f(\{u\}\cup S| \psi_{[t]})-f(S| \psi_{[t]})$ denotes the expected marginal benefit of $v$ given existing seeds $S$ and partial realization $\psi_{[t]}$ (i.e., we select a node that maximizes the expected marginal benefit).
 Otherwise, if condition (\ref{cond:!}) is not satisfied, we wait for another slot.
\item This process iterates until  the budget is used up.
\end{enumerate}

Condition (\ref{cond:!}) can be interpreted as follows: we will not select the next seed until the average activation probability of all nodes with non-zero activation probability is sufficiently high. Note that this condition can always be satisfied when we reach a final realization (i.e., $\frac{f(S| \psi_{[t]})}{|V\setminus O(S; \psi_{[t]})|}=1$ when $\psi_{[t]}$ is a final realization). 
We use $\alpha$ to control the tradeoff of \emph{delay} and \emph{performance}. In particular, a larger $\alpha$ indicates longer delay but better performance. For example, if we set $\alpha=1$, our model becomes full-feedback model, that is, we must wait until every node is either in active state or non-active state, before selecting the next seed. On the other hand, if we set $\alpha=0$, our model is reduced to zero-feedback model, implying that we can select all seeds in advance. 

 It was worth noting that we may select multiple seeds in one slot as long as condition (\ref{cond:!}) holds, thus one slot may contain multiple rounds.
\begin{algorithm}[hptb]
\caption{$\alpha$-Greedy Policy: $\pi^\alpha$}
\label{alg:LPP1}
\textbf{Input}: $0 \leq \alpha \leq 1$.\\
\textbf{Output}: $S$.
\begin{algorithmic}[1]
\STATE $S=\emptyset$; $t=0$;
\STATE $S\leftarrow S \cup\{\arg\max_{u \in V }\Delta_u (S| \psi_{[t]})\}$;
\STATE $B \leftarrow B-1$; $t \leftarrow t+1$
\WHILE {$B \geq 0$}
\IF {$\frac{f(S| \psi_{[t]})}{|V\setminus O(S; \psi_{[t]})|}\geq \alpha
$}
\STATE $S\leftarrow S \cup \{\arg\max_{u \in V}\Delta_u (S| \psi_{[t]})\}$;
\STATE $B \leftarrow B-1$; 
\ELSE
\STATE $t \leftarrow t+1$; update $\psi_{[t]}$;
\ENDIF
\ENDWHILE
\RETURN $S$
\end{algorithmic}
\end{algorithm}


\subsubsection{Performance Analysis} Let $\pi_{[t]}$ denote the level-$t$-truncation of
 $\pi$ obtained by running until it terminates or until slot $t$.   For every $1 \leq i\leq B$, assume the $i$-th seed is added to $S$ at slot $t_i$ (i.e., $\frac{f(S| \psi_{[t_i]})}{|V\setminus O(S; \psi_{[t_i]})|}\geq \alpha$).
 For brevity, we use $O$ to denote $O(S; \psi_{[t_i]})$. Let $f^{O}(S| \psi_{[t_i]})=\sum_{v\in O} p_v(S; \psi_{[t_i]})$ denote the expected cascade in ${O}$ given seeds $S$ and observation $\psi_{[t_i]}$. We use $\pi^*$ to denote the optimal adaptive policy. In the rest of this paper, let $f(\pi| \psi)$,  $ f^O(\pi| \psi)$, $f^{V\setminus O}(\pi| \psi)$  denote respectively the expected cascade of $\pi$ in $V$, $O$, $V\setminus O$ under  realization $\psi$. In order to prove the main theorem (Theorem \ref{thm:main}),  we first prove two preparatory lemmas (Lemma \ref{lem:1} and Lemma \ref{lem:2}).

\begin{lemma}
\label{lem:1}
For any $1 \leq i \leq B$ and $ 0<\alpha\leq 1$, we have
\[f(\pi^*|\psi_{[t_i]}) - f(\pi^\alpha_{[t_i]}|\psi_{[t_i]})/\alpha \leq f^{O}(\pi^*|\psi_{[t_i]}) - f^{O}(\pi^\alpha_{[t_i]}|\psi_{[t_i]})\]
\end{lemma}
\emph{Proof:}  First, we have
\begin{align}
&f(\pi^*|\psi_{[r_i]}) - f(\pi^\alpha_{[t_i]}|\psi_{[t_i]})/\alpha \\
&=  (f^{O}(\pi^*|\psi_{[t_i]}) - f^{O}(\pi^\alpha_{[t_i]}|\psi_{[t_i]})/\alpha) +  (f^{V\setminus O}(\pi^*|\psi_{[t_i]}) - f^{V\setminus O}(\pi^\alpha_{[t_i]}|\psi_{[t_i]})/\alpha) \\
& \leq (f^{O}(\pi^*|\psi_{[t_i]}) - f^{O}(\pi^\alpha_{[t_i]}|\psi_{[t_i]})) +  (f^{V\setminus O}(\pi^*|\psi_{[t_i]}) - f^{V\setminus O}(\pi^\alpha_{[t_i]}|\psi_{[t_i]})/\alpha)
\end{align}
Based on the definition of $t_i$, we have $\frac{\sum_{v\in V}p_v(S; \psi_{[t_i]})}{|V\setminus O|} = \frac{\sum_{v\in V\setminus O}p_v(S; \psi_{[t_i]})}{|V\setminus O|}\geq \alpha$. It follows that $f^{V\setminus O}(\pi^\alpha_{[t_i]}|\psi_{[t_i]})= \sum_{v\in V\setminus O}p_v(S; \psi_{[t_i]})\geq \alpha |V\setminus O|$, thus $f^{V\setminus O}(\pi^*|\psi_{[t_i]}) - f^{V\setminus O}(\pi^\alpha_{[t_i]}|\psi_{[t_i]})/\alpha \leq 0$. Then
$f(\pi^*|\psi_{[t_i]}) - f(\pi^\alpha_{[t_i]}|\psi_{[t_i]})/\alpha \leq f^{O}(\pi^*|\psi_{[t_i]}) - f^{O}(\pi^\alpha_{[t_i]}|\psi_{[t_i]})$. $\Box$


\begin{lemma}
\label{lem:2}
For any $1 \leq i \leq B$,  we have
\[\max_{u\in V}\Delta_u (S; \psi_{[t_i]}) \geq \frac{1}{B} \left (f^{O}(\pi^*|\psi_{[t_i]}) - f^{O}(\pi^\alpha_{[t_i]}|\psi_{[t_i]})\right)\]
\end{lemma}
\emph{Proof:} Assume $\psi' $ and $\psi$ are two partial realizations satisfying $\psi' \supseteq \psi$, let $S'$ and $S$ denote the seeds selected under $\psi' $ and $\psi$, satisfying $S'\supseteq S$. Since the activation probability of $v\in O(S; \psi)$ is zero, then based on similar proof of Theorem 8.1 in \citep{golovin2011adaptive}, we can prove the following
$\forall v\in O(S;\psi), \forall z\in V: p_v(S\cup\{z\}; \psi) - p_v(S; \psi)\geq p_v(S'\cup\{z\}; \psi') - p_v(S'; \psi') $.

Notice that $f^{O(S; \psi)}(S| \psi) = \sum_{v\in O(S; \psi)} p_v(S; \psi)$, it follows that $f^{O(S; \psi)}(S\cup\{z\}|\psi)-f^{O(S; \psi)}(S| \psi) \geq  f^{O(S; \psi)}(S'\cup\{z\}| \psi')-f^{O(S; \psi)}(S'| \psi')$. This implies that  function $f^O(\cdot)$ is submodular. Based on  the standard analysis of submodular maximization, we have $\max_{z\in V} [f^O(S\cup\{z\}| \psi_{[t_i]})-f^O(S| \psi_{[t_i]})] \geq \frac{1}{B} \left (f^{O}(\pi^*|\psi_{[t_i]}) - f^{O}(\pi^\alpha_{[t_i]}|\psi_{[t_i]})\right)$. Because $\max_{u\in V}\Delta_u (S; \psi) \geq \max_{z\in V} [f^O(S\cup\{z\}| \psi_{[t_i]})-f^O(S| \psi_{[t_i]})]$, we have $\max_{u\in V}\Delta_u (S; \psi) \geq \frac{1}{B} \left (f^{O}(\pi^*|\psi_{[t_i]}) - f^{O}(\pi^\alpha_{[t_i]}|\psi_{[t_i]})\right)$.
$\Box$

Now we are ready to prove the performance bound of $\pi^\alpha$.
\begin{theorem}
\label{thm:main}
The expected cascade of $\pi^\alpha$  is bounded by
$f(\pi^\alpha) \geq \alpha (1-e^{-\frac{1}{\alpha}})  f(\pi^*)$.
\end{theorem}
\emph{Proof:} Let $\Delta_i = f(\pi^*) - f(\pi^\alpha_{[t_i]})/\alpha$ and $\Pr(\psi_{[t_{i-1}]})$ denote the probability that $\psi_{[t_{i-1}]}$ is observed at slot $t_{i-1}$, we have
\begin{align}\alpha (\Delta_{i-1}-\Delta_{i})
&= f(\pi^\alpha_{[t_{i}]}) - f(\pi^\alpha_{[t_{i-1}]}) =\Pr(\psi_{[t_{i-1}]}) (f(\pi^\alpha_{[t_{i}]}|\psi_{[t_{i-1}]})-f(\pi^\alpha_{[t_{i-1}]}|\psi_{[t_{i-1}]})) \\
&\geq \Pr(\psi_{[t_{i-1}]})(\frac{1}{B} \left (f^{O}(\pi^*|\psi_{[t_{i-1}]}) - f^{O}(\pi^\alpha_{[t_{i-1}]}|\psi_{[t_{i-1}]})\right))\\
&\geq \Pr(\psi_{[t_{i-1}]})(\frac{1}{B} (f(\pi^*|\psi_{[t_{i-1}]}) - f(\pi^\alpha_{[t_{i-1}]}|\psi_{[t_{i-1}]})/\alpha)) \\
&= \frac{1}{B} f(\pi^*)- \frac{1}{\alpha B}\Pr(\psi_{[t_{i-1}]})f(\pi^\alpha_{[t_{i-1}]}|\psi_{[t_{i-1}]})\\
&= \frac{1}{B} f(\pi^*)- \frac{1}{\alpha B}f(\pi^\alpha_{[t_i-1]}) = \frac{1}{B} \Delta_{i-1}
\end{align}
The first inequality is due to Lemma \ref{lem:2}, and the second inequality is due to Lemma \ref{lem:1}.
It follows that $\Delta_{i+1}  \leq (1-\frac{1}{\alpha B})\Delta_i$.
Hence $\Delta_B  \leq (1-\frac{1}{\alpha B})^B\Delta_0 \leq e^{-\frac{1}{\alpha}} \Delta_0$.
It follows that
$f(\pi^*) - f(\pi^\alpha_{[t_B]})/\alpha \leq e^{-\frac{1}{\alpha}} \Delta_0 = e^{-\frac{1}{\alpha}} f(\pi^*)$.
Hence
$f(\pi^\alpha)= f(\pi^\alpha_{[t_B]}) \geq \alpha (1-e^{-\frac{1}{\alpha}})  f(\pi^*)$.
$\Box$


As a corollary of Theorem \ref{thm:main}, we can prove that the approximation ratio of our greedy policy under full-feedback setting is $(1-e^{-1})$.

\begin{corollary} Under full-feedback model, i.e., $\alpha=1$, we have
$f(\pi^\alpha) \geq (1-e^{-1})  f(\pi^*)$.
\end{corollary}

Another interesting finding is that if we set $\alpha=0$, then  condition (\ref{cond:!}) is always true regardless of the observation $\psi_{[t]}$.
Our policy under this setting becomes \emph{non-adaptive}, implying that we can select all $B$ seeds in advance without observing any partial realization. It is also worth noting that select the seed at a later slot never worsens the result. 

\textbf{Implications of our results.} One immediate implication of Theorem \ref{thm:main} is that given a desired approximation ratio, we can decide the appropriate seed selection point.  Another implication is that we can decide the approximation ratio of the greedy selection strategy given any partial feedback.

\subsection{Non-Uniform Cost}
\subsubsection{Policy Description}
We next study the case with non-uniform cost. The previous adaptive policy can be naturally modified to handle non-uniform item costs by replacing its
selection rule by selecting $\arg\max_{u \in V}\frac{\Delta_u (S; \psi_{[t]})}{c_u}$ (i.e.,  select the node that achieves the largest \emph{benefit-to-cost} ratio).
The detailed description of  our greedy policy with non-uniform cost is listed in Algorithm \ref{alg:LPP22}.

\begin{algorithm}[hptb]
\caption{$\alpha$-Greedy Policy with non-uniform cost: $\pi^{\alpha}$}
\label{alg:LPP22}
\textbf{Input}: $0 \leq \alpha \leq 1$.\\
\textbf{Output}: $S$.
\begin{algorithmic}[1]
\STATE $S=\emptyset$; $t=0$;
\STATE select $v=\arg\max_{u \in V }\frac{\Delta_u (S; \psi_{[t]})}{c_u}$;
\STATE $S\leftarrow S \cup \{v\}$; $B \leftarrow B-c_v$;
\STATE $t \leftarrow t+1$;  update $\psi_{[t]}$;
\WHILE {$B \geq 0$}
\IF {$\frac{f(S| \psi_{[t]})}{|V\setminus O(S;\psi_{[t]})|}\geq \alpha$}
\STATE select $v=\arg\max_{u \in V}\frac{\Delta_u (S; \psi_{[t]})}{c_u}$;
\IF {$B-c_v < 0$}
\STATE break;
\ELSE
\STATE $S\leftarrow S \cup \{v\}$; $B \leftarrow B-c_v$;
\ENDIF
\ELSE
\STATE $t \leftarrow t+1$;
\STATE  update $\psi_{[t]}$;
\ENDIF
\ENDWHILE
\RETURN $S$
\end{algorithmic}
\end{algorithm}

\subsubsection{Performance Analysis}
\label{sec:alpha}
We next focus on analyzing the performance bound of $\pi^\alpha$. For this purpose, we introduce the concepts of \emph{virtual time} and \emph{virtual slot}. Imagine that a policy $\pi$ runs over virtual time, after $\pi$ adds a node $u$ to $S$, it starts to run $u$, and stops after $c_u$ virtual slots. Notice that running a node over virtual time does not consume actual time. We next introduce two important concepts. The level-$k$-truncation of a policy $\pi$  over virtual time, denoted by $\pi_k$, is a
randomized policy defined as follows.

 \begin{definition}[Level-$k$-truncation of $\pi$ over virtual time] First run $\pi$ for $k$ virtual slots, and for every node $v\in V$, if  $v$ has been running for $\gamma\leq c_v$ virtual slots, selecting $v$ independently with probability $\gamma/c_v$. 
 \end{definition}

The strict level-$k$-truncation of $\pi$, denoted  $\pi_{\leftarrow k}$, is defined as follows.
 \begin{definition}[Strict level-$k$-truncation of $\pi$  over virtual time] First run $\pi$ for $k$ virtual slots, and for every node $v\in V$, selecting $v$ if and only if it has been run for $\gamma= c_v$ virtual slots (i.e., removing any node whose virtual running time is strictly less than its cost).
 \end{definition}

\begin{theorem}
\label{thm:2}
Let $\overline{c}=\max_{v\in V} c_v$, the expected cascade of $\pi^\alpha$ is bounded by
$f(\pi^\alpha) \geq \alpha (1-e^{-\frac{1}{\alpha} \frac{B-\overline{c}}{B}})  f(\pi^*)$.
\end{theorem}
\emph{Proof:} Our proof is based on techniques studied in the context of adaptive submodular maximization \cite{golovin2011adaptive}. In the rest of the proof, let $S$ denote the set of  seeds selected by $\pi^\alpha_{\leftarrow k}$, $O$ denote the set of nodes whose activation probability is zero under $\psi_{k}$ and $S$, and $\psi_{k}$ denote the partial realization observed at virtual slot $k$. We first provide two preparatory lemmas whose proofs are similar to the proofs of Lemma \ref{lem:1} and Lemma \ref{lem:2}.
\begin{lemma}
\label{lem:111}
For any $k \geq 1$, we have
$f(\pi^*|\psi_{k}) - f(\pi^\alpha_{ \leftarrow k}|\psi_{k})/\alpha \leq f^{O}(\pi^*|\psi_{k}) - f^{O}(\pi^\alpha_{\leftarrow k}|\psi_{k})$.
\end{lemma}

\begin{lemma}
\label{lem:222}
For any $k \geq 1$, $\max_{u\in V}(\Delta_u (S; \psi_{k})/c_u) \geq\frac{1}{B} \left (f^{O}(\pi^*|\psi_{k}) - f^{O}(\pi^\alpha_{\leftarrow k}|\psi_{k})\right)$.
\end{lemma}

Let $\Delta_k = f(\pi^*) - f(\pi_{k}^{\alpha})/\alpha$ and $\Pr[\psi_{i}]$ denote the probability that $\psi_{i}$ is observed in virtual slot $i$, we have
\begin{align*}
\alpha (\Delta_{k-1}-\Delta_{k})&= f(\pi^\alpha_{i})-f(\pi^\alpha_{k-1}) \\
&=\Pr[\psi_{i-1}] \max_{v\in V}(\Delta_v (S; \psi_{k-1})/c_v) \\
&\geq  \frac{1}{B}\Pr[\psi_{k-1}] (f^{O}(\pi^*|\psi_{k-1}) - f^{O}(\pi^\alpha_{\leftarrow (k-1)}|\psi_{k-1}))\\
&\geq    \frac{1}{B} \Pr[\psi_{k-1}] (f(\pi^*|\psi_{k-1}) - f(\pi^\alpha_{\leftarrow (k-1)}|\psi_{k-1})/\alpha)) \\
&\geq    \frac{1}{B} \Pr[\psi_{k-1}] (f(\pi^*|\psi_{k-1}) - f(\pi^\alpha_{ k-1}|\psi_{k-1})/\alpha)) \\
&=     \frac{1}{B} \Delta_{k-1}
\end{align*}
 The first inequality is due to Lemma \ref{lem:222} and the second inequality is due to Lemma \ref{lem:111}. It follows that
$\Delta_{k}  \leq (1-  \frac{1 }{\alpha B})\Delta_{k-1}$.
Hence, assume $\pi^\alpha$ terminates after $l$ virtual slots, we have
$\Delta_l  \leq [\prod_{k=1}^l (1- \frac{1}{\alpha B})] \Delta_0 \leq e^{-\frac{1}{\alpha} \frac{l}{B}} \Delta_0$ where for the second inequality we have used the fact that $1-x < e^{-x}$ for all $x>0$. Hence
\begin{equation}
\label{eq:important}
f(\pi^\alpha_{l}) \geq \alpha (1-e^{-\frac{1}{\alpha} \frac{l}{B}})  f(\pi^*)
\end{equation}

Although we are not guaranteed to  use up all the budget at the end of $\pi^\alpha$, the remaining budget in every realization can not be larger than  $\overline{c}$. In other words, the last virtual slot reached by $\pi_\alpha$ is $l\geq B-\overline{c}$.
It follows that
$f(\pi^\alpha) \geq \pi^\alpha_{B-\overline{c}}\geq\alpha (1-e^{-\frac{1}{\alpha} \frac{B-\overline{c}}{B}})  f(\pi^*)$. $\Box$

Based on $\pi^{\alpha}$, we next provide an enhanced greedy policy $\pi^{\alpha-\mathrm{enhanced}}$ with constant approximation ratio. $\pi^{\alpha-\mathrm{enhanced}}$ (Algorithm \ref{alg:LPP2}) randomly picks one from the following two candidate solutions with equal probability: The first candidate solution contains a single node $v^*$ which can maximize the expected cascade: $v^* = \arg\max_{v\in V}I(\{v\})$; the second candidate solution is computed by the greedy policy $\pi^{\alpha}$.  

\begin{algorithm}[hptb]
\caption{Enhanced Greedy Policy $\pi^{\alpha-\mathrm{enhanced}}$}
\label{alg:LPP2}
\begin{algorithmic}[1]
\STATE Let $v^* = \arg\max_{v\in V}I(\{v\})$;
\STATE Randomly pick one from the following two strategies with equal probability: return $\{v^*\}$ or run $\pi^{\mathrm{\alpha}}$;
\end{algorithmic}
\end{algorithm}

\begin{theorem}
\label{thm:main3}
The expected cascade of $\pi^{\alpha-\mathrm{enhanced}}$ is bounded by
$f(\pi^{\alpha-\mathrm{enhanced}}) \geq \frac{\alpha (1-e^{-\frac{1}{\alpha}})}{2}  f(\pi^*)$.
\end{theorem}
\emph{Proof:} Consider a policy $\pi'$ obtained by first running $\pi^\alpha$, then selecting one more node according to the same greedy manner. It is easy to verify that $\pi'$ runs for at least $B$ virtual slots. According to Theorem \ref{thm:2}, we have $f(\pi') \geq \alpha (1-e^{-\frac{1}{\alpha}})  f(\pi^*)$.  Due to the submodularity of $I(\cdot)$,  we have $f(\pi') \leq f(\pi^{\alpha}) + I(\{v^*\})$. Thus
$f(\pi^{\alpha}) + I(\{v^*\})  \geq f(\pi') \geq \alpha (1-e^{-\frac{1}{\alpha}}) f(\pi^*)$.
Then $f(\pi^{\alpha-\mathrm{enhanced}}) = (f(\pi^{\alpha}) + I(\{v^*\}) )/2 \geq  \frac{\alpha (1-e^{-\frac{1}{\alpha}})}{2}  f(\pi^*)$.

%
%
%
$\Box$

As a corollary of Theorem \ref{thm:main3}, we can prove that the approximation ratio of $\pi^{\alpha-\mathrm{enhanced}}$  under full-feedback and non-uniform cost setting is $(1-e^{-1})/2$.
\begin{corollary} Under full-feedback model, i.e., $\alpha=1$, we have
$f(\pi^{\alpha-\mathrm{enhanced}}) \geq \frac{(1-e^{-1})}{2}  f(\pi^*)$.
\end{corollary}

%

\section{$\beta$-Greedy  Adaptive Policy with Partial Feedback}
 Our second policy $\pi^\beta$ (Algorithm \ref{alg:LPP11}) is still a simple greedy policy. However, we replace condition (\ref{cond:!}) used in the previous policy by condition (\ref{cond:2}). We first explain the idea of $\beta$-Greedy policy with non-uniform cost and then analyze the performance bound of $\pi^\beta$.
\subsection{Policy Description}
We next summarize the work flow of $\beta$-Greedy policy with non-uniform cost.
\begin{enumerate}
\item Start with slot $t=1$, $S=\emptyset$, and a control parameter $\beta\in[0,1]$;
\item Suppose we have made observations $\psi_{[t]}$ at slot $t$,  let $\Pr[\psi|\psi_{[t]}]$ denote the probability that $\psi$ is the final superrealization of $\psi_{[t]}$. Given $S$ and final realization $\psi$,  let  $\max_{u\in V}(\Delta_u (S; \psi)/c_u)$ denote the largest benefit-to-cost ratio  achieved by selecting one more node.   We then examine the following condition.
\begin{equation}
\label{cond:2}
\frac{\max_{v\in V} (\Delta_v (S; \psi_{[t]})/c_v)}{E_{\Psi}[\max_{u\in V}(\Delta_u (S; \Psi)/c_u)|\Psi\in X_{\mathrm{final}}(S|\psi_{[t]})]} \geq \beta
\end{equation}
where
    \[E_{\Psi}[\max_{u\in V}(\Delta_u (S; \Psi)/c_u)|\Psi\in X_{\mathrm{final}}(S|\psi_{[t]})]=\sum_{\psi\in X_{\mathrm{final}}(S|\psi_{[t]})}\Pr[\psi|\psi_{[t]}]\max_{u\in V}(\Delta_u (S; \psi)/c_u)\]
 If  condition (\ref{cond:2}) is satisfied,  add $\arg\max_{u \in V }(\Delta_u(S; \psi_{[t]})/c_u)$ to $S$ (i.e., we select a node that maximizes the expected benefit-to-cost ratio).
 Otherwise, if condition (\ref{cond:2}) is not satisfied, wait for another slot.
\item This process iterates until  the budget is used up.
\end{enumerate}

The left side of  condition (\ref{cond:2}) can be interpreted as the gap between  adaptive policy and non-adaptive policy given $\psi_{[t]}$. 
We use $\beta$  to control the tradeoff of delay and performance. It was worth noting that  $E_{\Psi}[\max_{u\in V}(\Delta_u (S; \Psi)/c_u)|\Psi\in X_{\mathrm{final}}(S|\psi_{[t]})]$ in  condition (\ref{cond:2}) can be replaced by $\max_{\psi\in X_{\mathrm{final}}(S|\psi_{[t]})}\max_{u\in V}(\Delta_u (S; \psi)/c_u)$ without affecting our results, this is due to $\max_{\psi\in X_{\mathrm{final}}(S|\psi_{[t]})}\max_{u\in V}(\Delta_u (S; \psi)/c_u) \geq E_{\Psi}[\max_{u\in V}(\Delta_u (S; \Psi)/c_u)|\Psi\in X_{\mathrm{final}}(S|\psi_{[t]})]$.  However, this may  prolong the waiting time before selecting the next seed. In fact, \cite{tang2018social} uses this value to derive a performance bound on adaptive influence maximization problem subject to partition matroid constraint.  As pointed out in \citep{tang2018social}, $\arg\max_{\psi\in X_{\mathrm{final}}(S|\psi_{[t]})}\max_{u\in V}(\Delta_u (S; \psi)/c_u)$ can be interpreted as the most ``pessimistic'' final superrealization of $\psi_{[t]}$ under which no additional users, other than those who have been influenced under $\psi_{[t]}$, will be influenced. As compared with $E_{\Psi}[\max_{u\in V}(\Delta_u (S; \Psi)/c_u)|\Psi\in X_{\mathrm{final}}(S|\psi_{[t]})]$, it is easier to evaluate the value of $\max_{\psi\in X_{\mathrm{final}}(S|\psi_{[t]})}\max_{u\in V}(\Delta_u (S; \psi)/c_u)$. Assume that no additional nodes can be further influenced given $\psi_{[t]}$, we select the node, say $v$, that maximizes the marginal benefit-to-cost ratio. It is easy to verify that the benefit-to-cost ratio of $v$ is $\max_{\psi\in X_{\mathrm{final}}(S|\psi_{[t]})}\max_{u\in V}(\Delta_u (S; \psi)/c_u)$.

\begin{algorithm}[hptb]
\caption{$\beta$-Greedy Policy: $\pi^\beta$}
\label{alg:LPP11}
\textbf{Input}: $0 \leq \beta \leq 1$.\\
\textbf{Output}: $S$.
\begin{algorithmic}[1]
\STATE $S=\emptyset$; $t=0$;
\STATE select $v=\arg\max_{u \in V }\frac{\Delta_u (S; \psi_{[t]})}{c_u}$;
\STATE $S\leftarrow S \cup \{v\}$; $B \leftarrow B-c_v$;
\STATE $t \leftarrow t+1$;  update $\psi_{[t]}$;
\WHILE {$B \geq 0$}
\IF {$\frac{\max_{v\in V} (\Delta_v (S; \psi_{[t]})/c_v)}{E_{\Psi}[\max_{u\in V}(\Delta_u (S; \Psi)/c_u)|\Psi\in X_{\mathrm{final}}(S|\psi_{[t]})]} \geq \beta
$}
\STATE select $v=\arg\max_{u \in V}\frac{\Delta_u (S; \psi_{[t]})}{c_u}$; \label{line:1}
\IF {$B-c_v < 0$}
\STATE break;
\ELSE
\STATE $S\leftarrow S \cup \{v\}$; $B \leftarrow B-c_v$;
\ENDIF
\ELSE
\STATE $t \leftarrow t+1$; \STATE  update $\psi_{[t]}$; 
\ENDIF
\ENDWHILE
\RETURN $S$
\end{algorithmic}
\end{algorithm}

\subsection{Performance Analysis} 
We next provide the performance bound of $\pi^\beta$. In the rest of this section, we adopt the same notations used in Section \ref{sec:alpha}.

\begin{theorem}
\label{thm:7}
Let $\overline{c}=\max_{v\in V} c_v$, the expected cascade of $\pi^\beta$ is bounded by
$f(\pi^\beta) \geq  (1-e^{- \beta \frac{B-\overline{c}}{B}})  f(\pi^*)$.
\end{theorem}
\emph{Proof:}  Let $\Delta_k = f(\pi^*) - f(\pi^\beta_{k})$, we have
\begin{align*}\Delta_{k-1}-\Delta_{k}
&= f(\pi^\beta_{{k}}) - f(\pi^\beta_{{k-1}}) =\Pr[\psi_{k-1}](f(\pi^\beta_{k}|\psi_{k-1})-f(\pi^\beta_{k-1}|\psi_{k-1})) \\
&\geq \Pr[\psi_{k-1}]\max_{v\in V}(\Delta_v (S; \psi_{k-1})/c_v) \\
&\geq \Pr[\psi_{k-1}](\beta\sum_{\psi\in X_{\mathrm{final}}(S|\psi_{k-1})}\Pr[\psi|\psi_{k-1}]\max_{u\in V}(\Delta_u (S; \psi)/c_u))\\
&\geq \Pr[\psi_{k-1}]\frac{\beta}{B}(f(\pi^\beta_{[\leftarrow(k-1)]}@\pi^*|\psi_{k-1})-f(\pi^\beta_{[\leftarrow(k-1)]}|\psi_{k-1}))\\
&\geq \Pr[\psi_{k-1}]\frac{\beta}{B}(f(\pi^*|\psi_{k-1})-f(\pi^\beta_{k-1}|\psi_{k-1})) \label{eq:111
}\\
&= \frac{\beta}{B} (f(\pi^*)- f(\pi^\beta_{k-1})) \geq  \frac{\beta}{B} \Delta_{k-1}
\end{align*}
The first inequality is due to the selection rule of  $\pi^\beta_k$ (Line \ref{line:1} in Algorithm \ref{alg:LPP11}). The second inequality holds due to condition (\ref{cond:2}) is satisfied before selecting the next node. The third inequality is due to the adaptive submodularity of $f$ under full-feedback model. The last inequality is due to $f(\pi^*)\geq f(\pi^\beta_{[\leftarrow(k-1)]}@\pi^*)$ and $f(\pi^\beta_{[\leftarrow(k-1)]})\geq (\pi^\beta_{k-1})$.
It follows that $\Delta_{k}  \leq (1-\frac{\beta}{ B})\Delta_{k-1}$.
Hence $\Delta_l  \leq (1-\frac{\beta}{ B})^l\Delta_0 \leq e^{-\frac{\beta l}{B}} \Delta_0$.
It follows that
$f(\pi^*) - f(\pi^\beta_{l}) \leq e^{-\beta} \Delta_0 = e^{-\frac{\beta l}{B}} f(\pi^*)$. Because the last virtual slot reached by $\pi_\beta$ is $l\geq B-\overline{c}$, we have $f(\pi^\beta)\geq f(\pi^\beta_{B-\overline{c}}) \geq  (1-e^{-\frac{\beta (B-\overline{c})}{B}})  f(\pi^*)$.
$\Box$


As a corollary of Theorem \ref{thm:main}, we can prove that the approximation ratio of our greedy policy under full-feedback setting is $(1-e^{-1})$.

\begin{corollary} Under full-feedback model, i.e., $\beta=1$, we have
$f(\pi^\beta) \geq (1-e^{-1})  f(\pi^*)$.
\end{corollary}
%

Based on $\pi^{\beta}$, we next provide an enhanced greedy policy $\pi^{\beta-\mathrm{enhanced}}$ (Algorithm \ref{alg:LPP2}) with constant approximation ratio. $\pi^{\beta-\mathrm{enhanced}}$  randomly picks one from the following two candidate solutions with equal probability: The first candidate solution contains a single node $v^*$ which can maximize the expected cascade: $v^* = \arg\max_{v\in V}I(\{v\})$; the second candidate solution is computed by the greedy policy $\pi^{\beta}$.  

\begin{algorithm}[hptb]
\caption{Enhanced Greedy Policy $\pi^{\beta-\mathrm{enhanced}}$}
\label{alg:LPP2}
\begin{algorithmic}[1]
\STATE Let $v^* = \arg\max_{v\in V}I(\{v\})$;
\STATE Randomly pick one from the following two strategies with equal probability: return $\{v^*\}$ or run $\pi^{\beta}$;
\end{algorithmic}
\end{algorithm}

\begin{theorem}
\label{thm:main1}
The expected cascade of $\pi^{\beta-\mathrm{enhanced}}$ is bounded by
$f(\pi^{\beta-\mathrm{enhanced}}) \geq \frac{(1-e^{-\beta})  }{2}  f(\pi^*)$.
\end{theorem}
\emph{Proof:}  Due to the submodularity of $I(\cdot)$,  we have
$f(\pi^{\mathrm{\beta}}) + I(\{v^*\})  \geq (1-e^{-\beta}) f(\pi^*)$.
Then $f(\pi^{\beta-\mathrm{enhanced}}) = (f(\pi^{\beta}) + I(\{v^*\}) )/2 \geq  \frac{(1-e^{-\beta})  }{2}  f(\pi^*)$.

%
%
%
$\Box$

As a corollary of Theorem \ref{thm:main1}, we can prove that the approximation ratio of $\pi^{\beta-\mathrm{enhanced}}$  under full-feedback and non-uniform cost setting is $(1-e^{-1})/2$.
\begin{corollary} Under full-feedback model, i.e., $\beta=1$, we have
$f(\pi^{\beta-\mathrm{enhanced}}) \geq \frac{(1-e^{-1})}{2}  f(\pi^*)$.
\end{corollary}

%

\section{Experimental Evaluation}

\begin{figure*}[hptb]
\hspace*{-0.5in}
\includegraphics[scale=0.2]{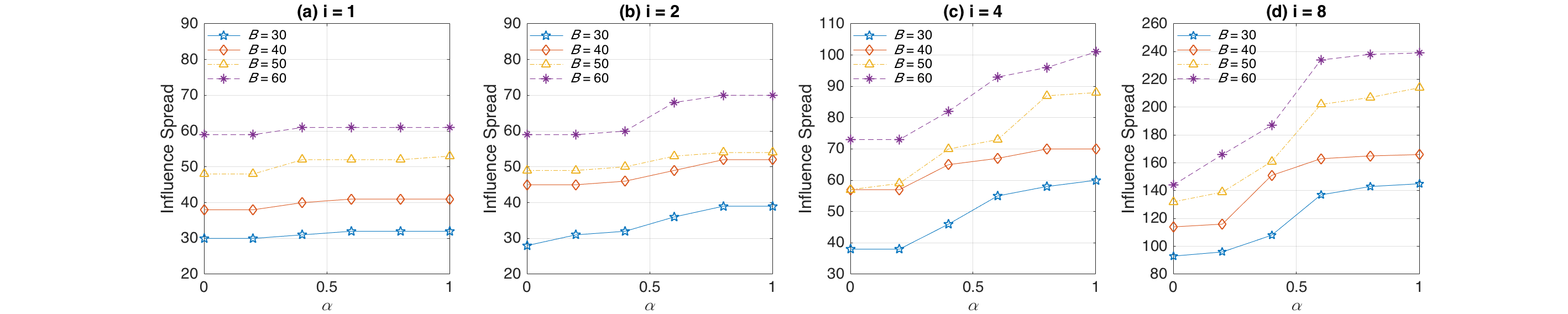}
\vspace{-0.3in}
\caption{Influence spread vs. $\alpha$ on \emph{NetHEPT} dataset under varying budget constraint.}
\label{fig:nethep}
\end{figure*}
We conduct extensive experiments on a real benchmark social networks: \emph{NetHEPT} to examine the effectiveness and efficiency of $\alpha$-Greedy policy. We set the propagation probability of each directed edge randomly from $i\times \{0.01, 0.001\}$ as in \citep{jung2012irie}. We vary the value of $i$ and examine how it affects the quality of the solutions. Selecting the node with the largest marginal gain is \#P-hard \citep{chen2010scalable}, and is typically approximated by numerous Monte Carlo simulations \citep{kempe2003maximizing}. We adjust the value of control parameter $\alpha$ in range $[0, 1]$. 

Figure \ref{fig:nethep} shows the influence spread yielded by the proposed enhanced greedy policy on the \emph{NetHEPT} dataset, as $\alpha$ ranges from $0$ to $1$ with a step of $0.2$. The $x$-axis corresponds to the value of the control parameter $\alpha$ and the $y$-axis holds the size of the influence spread achieved. We test the scenario with varying edge propagation probability distributions as discussed above. In particular, each edge is randomly assigned a propagation probability from $i\times \{0.01, 0.001\}$. We adjust the value of $i$ from $1$ to $8$ and Figure \ref{fig:nethep}(a)-(d) shows the comparison of the influence spread under $i=1$, $i=2$, $i=4$, $i=8$, respectively. In this set of experiments, the budget $B$ ranges from $30$ to $60$. The cost of each node is randomly assigned from $[1, 10]$. As expected, a higher budget leads to a larger influence spread since the budget indirectly controls the number of seeds can be selected.

We observe that when $i$ takes a smaller value, take $i=1$ as an example, the advantage of performing adaptive seeding with partial feedback is not obvious since the improvement over influence spread does not increase much as $\alpha$ increases. The reason behind this is that a smaller $i$ indicates a lower probability for the edges to be alive, resulting in a lower uncertainty about the status of the edges. In this case observations gained from partial feedback may not help much since with high probability the estimation of influence spread based on sampling technique matches the real propagation. As shown in Figure \ref{fig:nethep}, the advantage of taking adaptive seeding based on partial feedback becomes obvious as $i$ increases. We observe that when $i\geq 2$, a much larger influence spread can be achieved based on partial feedback ($\alpha>0$) compared to zero-feedback scenario ($\alpha=0$). For example, when $i=4$ with budget of $50$, while the influence spread based on zero-feedback leads to a size of $57$, the spread achieves a size of $87$ based on partial feedback ($\alpha=0.8$), a $52.6\%$ increase.

We also observe that a smaller $\alpha$ can lead to a significant improvement on influence spread with a higher edge propagation probability. For example, as shown in Figure \ref{fig:nethep}, when $i=2$, a $10\%$ improvement can be achieved with $\alpha=0.6$. When $i=4$, a $20\%$ improvement can be achieved with $\alpha=0.4$. This implies that given a social graph with moderate edge propagation probability, it is worth to leverage the partial observation of diffusion realization, since adaptive seeding based on partial feedback leads to a significant improvement over the size of influence spread.

%

\section{Conclusion}
To the best of our knowledge, we are the first to systematically study the problem of influence maximization problem with partial feedback. Under independent cascade model, which is one of the most commonly used models  in literature, we present two greedy algorithms with bounded approximation ratios.

\bibliographystyle{ormsv080}
\bibliography{reference}

\begin{thebibliography}{15}
\expandafter\ifx\csname natexlab\endcsname\relax\def\natexlab#1{#1}\fi
\expandafter\ifx\csname url\endcsname\relax
  \def\url#1{{\tt #1}}\fi
\expandafter\ifx\csname urlprefix\endcsname\relax\def\urlprefix{URL }\fi
\expandafter\ifx\csname urlstyle\endcsname\relax
  \expandafter\ifx\csname doi\endcsname\relax
  \def\doi#1{doi:\discretionary{}{}{}#1}\fi \else
  \expandafter\ifx\csname doi\endcsname\relax
  \def\doi{doi:\discretionary{}{}{}\begingroup \urlstyle{rm}\Url}\fi \fi

\bibitem[{Badanidiyuru et~al.(2016)Badanidiyuru, Papadimitriou, Rubinstein,
  Seeman, and Singer}]{badanidiyuru2016locally}
Badanidiyuru, Ashwinkumar, Christos Papadimitriou, Aviad Rubinstein, Lior
  Seeman, Yaron Singer. 2016.
\newblock Locally adaptive optimization: adaptive seeding for monotone
  submodular functions.
\newblock {\it Proceedings of the Twenty-Seventh Annual ACM-SIAM Symposium on
  Discrete Algorithms\/}. SIAM, 414--429.

\bibitem[{Chen et~al.(2013)Chen, Lakshmanan, and
  Castillo}]{chen2013information}
Chen, Wei, Laks~VS Lakshmanan, Carlos Castillo. 2013.
\newblock Information and influence propagation in social networks.
\newblock {\it Synthesis Lectures on Data Management\/} {\bf 5}(4) 1--177.

\bibitem[{Chen et~al.(2010)Chen, Wang, and Wang}]{chen2010scalable}
Chen, Wei, Chi Wang, Yajun Wang. 2010.
\newblock Scalable influence maximization for prevalent viral marketing in
  large-scale social networks.
\newblock {\it Proceedings of the 16th ACM SIGKDD international conference on
  Knowledge discovery and data mining\/}. ACM, 1029--1038.

\bibitem[{Chen et~al.(2009)Chen, Wang, and Yang}]{chen2009efficient}
Chen, Wei, Yajun Wang, Siyu Yang. 2009.
\newblock Efficient influence maximization in social networks.
\newblock {\it Proceedings of the 15th ACM SIGKDD international conference on
  Knowledge discovery and data mining\/}. ACM, 199--208.

\bibitem[{Cohen et~al.(2014)Cohen, Delling, Pajor, and
  Werneck}]{cohen2014sketch}
Cohen, Edith, Daniel Delling, Thomas Pajor, Renato~F Werneck. 2014.
\newblock Sketch-based influence maximization and computation: Scaling up with
  guarantees.
\newblock {\it Proceedings of the 23rd ACM International Conference on
  Conference on Information and Knowledge Management\/}. ACM, 629--638.

\bibitem[{Domingos and Richardson(2001)}]{domingos2001mining}
Domingos, Pedro, Matt Richardson. 2001.
\newblock Mining the network value of customers.
\newblock {\it Proceedings of the seventh ACM SIGKDD international conference
  on Knowledge discovery and data mining\/}. ACM, 57--66.

\bibitem[{Golovin and Krause(2011)}]{golovin2011adaptive}
Golovin, Daniel, Andreas Krause. 2011.
\newblock Adaptive submodularity: Theory and applications in active learning
  and stochastic optimization.
\newblock {\it Journal of Artificial Intelligence Research\/}  427--486.

\bibitem[{Jung et~al.(2012)Jung, Heo, and Chen}]{jung2012irie}
Jung, Kyomin, Wooram Heo, Wei Chen. 2012.
\newblock Irie: Scalable and robust influence maximization in social networks.
\newblock {\it Data Mining (ICDM), 2012 IEEE 12th International Conference
  on\/}. IEEE, 918--923.

\bibitem[{Kempe et~al.(2003)Kempe, Kleinberg, and Tardos}]{kempe2003maximizing}
Kempe, David, Jon Kleinberg, {\'E}va Tardos. 2003.
\newblock Maximizing the spread of influence through a social network.
\newblock {\it Proceedings of the ninth ACM SIGKDD international conference on
  Knowledge discovery and data mining\/}. ACM, 137--146.

\bibitem[{Leskovec et~al.(2007)Leskovec, Krause, Guestrin, Faloutsos,
  VanBriesen, and Glance}]{leskovec2007cost}
Leskovec, Jure, Andreas Krause, Carlos Guestrin, Christos Faloutsos, Jeanne
  VanBriesen, Natalie Glance. 2007.
\newblock Cost-effective outbreak detection in networks.
\newblock {\it Proceedings of the 13th ACM SIGKDD international conference on
  Knowledge discovery and data mining\/}. ACM, 420--429.

\bibitem[{Tang(2018)}]{tang2018social}
Tang, Shaojie. 2018.
\newblock When social advertising meets viral marketing: Sequencing social
  advertisements for influence maximization.
\newblock {\it Thirty-Second AAAI Conference on Artificial Intelligence\/}.

\bibitem[{Tang and Yuan(2016)}]{tang2016optimizing}
Tang, Shaojie, Jing Yuan. 2016.
\newblock Optimizing ad allocation in social advertising.
\newblock {\it Proceedings of the 25th ACM International on Conference on
  Information and Knowledge Management\/}. ACM, 1383--1392.

\bibitem[{Tang et~al.(2011)Tang, Yuan, Mao, Li, Chen, and
  Dai}]{tang2011relationship}
Tang, Shaojie, Jing Yuan, Xufei Mao, Xiang-Yang Li, Wei Chen, Guojun Dai. 2011.
\newblock Relationship classification in large scale online social networks and
  its impact on information propagation.
\newblock {\it INFOCOM, 2011 Proceedings IEEE\/}. IEEE, 2291--2299.

\bibitem[{Tong et~al.(2016)Tong, Wu, Tang, and Du}]{tong2016adaptive}
Tong, Guangmo, Weili Wu, Shaojie Tang, Ding-Zhu Du. 2016.
\newblock Adaptive influence maximization in dynamic social networks.
\newblock {\it IEEE/ACM Transactions on Networking\/} .

\bibitem[{Yuan and Tang(2017)}]{yuan2017discount}
Yuan, Jing, Shaojie Tang. 2017.
\newblock Adaptive discount allocation in social networks.
\newblock {\it Proceedings of the eighteenth ACM international symposium on
  Mobile ad hoc networking and computing\/}. ACM.

\end{thebibliography}

\end{document}